\documentstyle[12pt,aaspp4,psfig]{article}
\begin{document}

\title{Ionization Structure in the 30 Doradus\\Nebula as seen with HST/WFPC-2}

\author{P.A. Scowen\altaffilmark{1}, J.J. Hester\altaffilmark{1}, 
R. Sankrit\altaffilmark{1}, J.S. Gallagher \altaffilmark{2}, 
G.E. Ballester\altaffilmark{3}, 
C.J. Burrows\altaffilmark{4}, 
J.T. Clarke\altaffilmark{3}, 
D. Crisp\altaffilmark{5}, 
R.W. Evans\altaffilmark{5}, 
R.E. Griffiths\altaffilmark{6}, 
J.G. Hoessel\altaffilmark{2}, 
J.A. Holtzman\altaffilmark{7}, 
J. Krist\altaffilmark{4}, 
J.R. Mould\altaffilmark{8},
K.R. Stapelfeldt\altaffilmark{5}, 
J.T. Trauger\altaffilmark{5}, 
A.M. Watson\altaffilmark{9} 
and J.A. Westphal\altaffilmark{10}}

\altaffiltext{1}{Department of Physics and Astronomy, Box 871504,
Arizona State University, Tyler Mall, Tempe, AZ 85287-1504}
\altaffiltext{2}{Department of Astronomy, University of Wisconsin --
Madison, 475 N. Charter St., Madison, WI 53706}
\altaffiltext{3}{Department of Atmospheric and Oceanic Sciences,
University of Michigan, 2455 Hayward, Ann Arbor, MI 48109}
\altaffiltext{4}{Space Telescope Science Institute, 3700 San Martin
Drive, Baltimore, MD 21218}
\altaffiltext{5}{Jet Propulsion Laboratory, 4800 Oak Grove Drive,
Pasadena, CA 91109}
\altaffiltext{6}{Dept. of Physics - Wean Hall, Carnegie Mellon University,
5000 Forbes Ave., Pittsburgh, PA 15213-3890}
\altaffiltext{7}{Box 30001 / Dept. 4500, New Mexico State University,
Las Cruces, NM 88003}
\altaffiltext{8}{Mount Stromlo and Siding Springs Observatories,
Australian National University, Weston Creek Post Office, ACT 2611,
AUSTRALIA}
\altaffiltext{9}{Instituto de Astronom{\'\i}a UNAM, J. J. Tablada 1006,
Col. Lomas de Santa Maria, 58090 Morelia, Michoacan, MEXICO}
\altaffiltext{10}{Division of Geological and Planetary Sciences,
California Institute of Technology, Pasadena, CA 91125}

\centerline{accepted for publication in {\it The Astronomical Journal}, July
1998}

\begin{abstract}

Using the Hubble Space Telescope and WFPC2 we have imaged the central 20pc of
the giant H~II region 30 Doradus nebula in three different emission lines.  The
images allow us to study the nebula with a physical resolution that is within a
factor of two of that typical of ground based observations of Galactic H~II
regions.  We present a gallery of interesting objects within the region
studied.  These include a tube blown by the wind of a high velocity star and a
discrete H~II region around an isolated B star.  This small isolated H~II
region appears to be in the midst of the champagne flow phase of its evolution.

Most of the emission within 30 Dor is confined to a thin zone located between
the hot interior of the nebula and surrounding dense molecular material.  This
zone appears to be directly analogous to the photoionized photoevaporative
flows that dominate emission from small, nearby H~II regions.  For example, a
column of material protruding from the cavity wall to the south of the main
cluster is found to be a direct analog to elephant trunks in M16.  Surface
brightness profiles across this structure are very similar to surface
brightness profiles taken at ground based resolution across the head of the
largest column in M16.  The dynamical effects of the photoevaporative flow can
be seen as well.  An arcuate feature located above this column and a similar
feature surrounding a second nearby column are interpreted as shocks where the
photoevaporative flow stagnates against the high temperature gas that fills the
majority of the nebula.  The ram pressure in the photoevaporative flow, derived
from thermal pressure at the surface of the column, is found to balance with
the pressure in the interior of the nebula derived from previous x-ray
observations.  

By analogy with the comparison of ground and HST images of M16 we infer that
the same sharply stratified structure seen in HST images of M16 almost
certainly underlies the observed structure in 30 Dor.  30 Doradus is a crucial
case because it allows us to bridge the gap between nearby H~II regions and the
giant H~II regions seen in distant galaxies.  The real significance of this
result is that it demonstrates that the physical understanding gained from
detailed study of photoevaporative interfaces in nearby H~II regions can be
applied directly to interpretation of giant H~II regions.  Stated another way,
interpretation of observations of giant H~II regions must account for the fact
that this emission arises not from expansive volumes of ionized gas, but
instead from highly localized and extremely sharply stratified physical
structures.

\end{abstract}


\section{Introduction}

30 Doradus is a giant ionized complex in the Large Magellanic Cloud (LMC),
located at a distance of 51.3 kpc (eg. Panagia et al 1991).  The nebula is
centered on a dense cluster of newly formed stars, the most dense component of
which is called R136.  The nebula itself is more than 180 parsecs across,
qualifying it as a smaller member of the elite class of nebulae termed Giant
Extragalactic H~II Regions (GEHR's).  If 30 Doradus was placed at the distance
of the Orion Nebula from the Earth, it would appear to be more than 20 degrees
across, and would fill more than 4\% of the night sky.

The central cluster is very dense and is comprised of several hundred OB stars
with a small number of W-R stars (Hunter et al 1995b).  The integrated
ultraviolet flux from this cluster is intense: more than fifty times that being
produced in the center of the Orion Nebula (Campbell et al 1992).  Radiation
from the cluster combined with strong stellar winds from the most massive stars
in the cluster has eroded a large cavity in the nearby molecular complex,
producing the nebula we see today.

Hunter et al 1995b showed that the majority of the stars in the cluster were
formed in a single star formation event more than 2-3 million years ago.  The
census performed yielded a ``head count" of more than 3000 stars with more than
300 OB stars capable of producing the intense UV radiation and stong stellar
winds responsible for forming and shaping the H~II regions we observe in
galaxies.  The level of star formation exhibited by the 30 Doradus region and
the neighboring LMC complex are the closest example of starburst-like star
formation.  As such we are getting a unique view of the star formation
environment in the middle of an ongoing starburst.

The average reddening along the line of sight to the Large Magellanic Cloud and
30 Doradus is very low (Panagia et al 1991). However, within several H~II
complexes in the LMC comparison between optical and radio measurements suggest
a large variation in the local reddening. Kennicutt \& Hodge 1986 found a
variation in these estimates between 0 and 1 magnitude in A$_{V}$.  Hunter et
al 1995b also found substantial variation in the reddening across the face of
the 30 Doradus nebula, and derived a mean estimate for this reddening of 1.4
magnitudes in A$_{V}$ at 555 nm, and 0.8 magnitudes at 814 nm.  For the
purposes of this paper we will adopt an extinction of 1 magnitude in A$_{V}$
for the emission lines we observe.

The 30 Doradus nebula plays a key role in our understanding of H~II regions. 
Nearby regions are close enough for the physical processes at work within the
nebula to be studied in detail.  The work by Hester et al 1996 (hereafter H96)
on M16, for example, shows that emission within the nebula arises predominantly
within a narrow region at the interface between the H~II region and the
molecular cloud.  They follow Hester 1991 in describing this thin region as a
photoionized photoevaporative flow.  However, an H~II region like M16 is tiny
in comparison with giant H~II regions, and no giant H~II regions are close
enough to allow the stratified ionization structure of the photoevaporative
flow to be studied directly.  30 Doradus alone offers an opportunity to
bootstrap the physical understanding of small nearby H~II regions into the
context of the giant regions seen in distant galaxies.

In this paper we present Hubble Space Telescope images of the ionization
structure we observe around the central cluster R136.  The wealth of spatial
information contained in these pictures is daunting to consider, but we attempt
to summarize the most telling points by selecting and presenting several
examples of distinct structures around the field of view that provide insight
into how the interface with the local gas and dust is evolving.  In \S 2 we
discuss the observations themselves and the general structure of the nebula, as
well as presenting full-field mosaics of the data.  In \S 3 we discuss the
conditions apparent in the ionized hydrogen along the walls of the H~II region
cavity, as well as comparing the ionization structure we observe with models we
derive from the H$\alpha$ surface brightness.

\section{Observations and Data Processing}

Observations of 30 Doradus were obtained on January 2, 1994 using the Wide
Field and Planetary Camera 2 (WFPC-2).  Details of the observations are
presented in Table 1.  Results of the stellar content of R136 derived from the
broad-band images in this set have been published by Hunter et al 1995b.  Here
we present narrow-band observations of [O~III] $\lambda$5007, [S~II]
$\lambda\lambda$6717+6731, and H$\alpha$ $\lambda$6563 emission from the nebula
itself.  Images in a relatively line-free continuum band (F547M) were also
obtained to distinguish between line emission and stellar reflection luminosity.

The data were processed in the standard way for WFPC-2 (Holtzman et al 1995),
including subtraction of DC offsets, bias frames, and the flat field
correction.  Time dependent hot pixels were subtracted using maps derived from
on-orbit dark frames from that general time frame.  Cosmic rays were removed
using standard anti-coincidence techniques for each pair of images.  The images
were mosaicked using the appropriate astrometric solution from Holtzman et al
1995.

Figures 1-4 presents the four datasets used in this paper as surface brightness
mosaics: Figure 1 being the H$\alpha$, Figure 2 the [O~III], Figure 3 the
[S~II], and Figure 4 the F547M.  The location of the WFPC-2 field is
superimposed on a CTIO R-band image of 30 Doradus in Figure 5.  Figure 6
presents the full field color image of the 30 Doradus nebula.  In producing the
color image the following steps were taken:  (i) all stars were masked off and
filled, (ii) the [S~II] image was included as red, the H$\alpha$ as green and
the [O~III] as blue, and (iii) the F170W image was used to replace the hottest
stars also in blue.  As presented the position angle of the field of view for
these data is 62$^{\circ}$ east of north.  One WFC pixel represents a linear
scale of $7.7\times10^{16}$cm or 0.025pc, at the assumed distance of 51.3 kpc
to the nebula.

\subsection{General Appearance of the Nebula}

Our field of view is at the very heart of the 30 Doradus Nebula and positions
the cluster R136 on the Planetary Camera.  As such our field extends to a
radius of 15-20pc around the center of the nebula, with WF2 to the SE, WF3 to
the SW and WF4 to the NW (cf. Figure 1).  The structure of the nebula is
dominated by the ionizing flux from the large number of Wolf-Rayet and early O
stars known to be in the cluster (Hunter et al 1995b).  Very close to the
center of the cluster there is very little nebulosity, except for a component
that can be seen between the cluster stars.  This emission could originate
either in front of the cluster or behind it, but it does not lie within the
interior of the cluster.  Intense UV and stellar winds have cleared the cluster
itself of dense gas.

Almost all of the nebulosity we observe is concentrated into bright arcs or
edges located along the walls of the cavity.  Bright rims such as these are
commonplace in photoionized nebulae in our own Galaxy.  They represent the
boundary between the ionized volume and the nearby neutral and molecular gas
from which the young cluster was born.  The presence of significant xray
emission within the interior of the nebula and the strongly limb-brightened
appearance of these edges suggests that the more diffuse nebulosity arises in
such interfaces, but instead seen more face-on.

At a radius of about 5pc there is a bright circular rim centered on the
cluster.  This marks the first major interface between the cluster radiation
and the nebula wall.  To the WSW from R136 this rim is very bright and is
littered with finger-like structures that all appear to point toward the
cluster.  Outside of this rim the brightness of the nebula dims significantly.

One very bright rim is located directly south of the cluster and stands out as
a bright ridge set against a relatively darker region.  As such we contend that
this ridge is directly exposed to the ionizing radiation from the central
cluster.

The nature of the photoionized interface at the boundary of a molecular cloud
has been discussed in the literature.  Heating by UV radiation incident on the
interface drives a photoevaporative flow away from the interface (Bertoldi
1989, and references therein).  This flow is then photoionized by the same
radiation responsible for driving it.  Such photoionized photoevaporative flows
are characterized by sharply stratified ionization structure (Hester 1991;
H96).  The overall thickness of the resulting emissive region is typically of
order 10$^{17}$ cm, while the observed scale for stratification at the boundary
of the cloud can be much less.

By using emission lines arising from species with different ionization states
it is possible to probe the structure of this region.  High ionization lines
such as [O~III] $\lambda 5007$ arise in the part of the flow furthest from the
interface where the ionization parameter is high and the radiation field
contains enough high energy photons to reach such ionization states.  Emission
from species such as [S~II] $\lambda\lambda 6717+6731$ are brightest beyond the
hydrogen ionization edge where the gas density is high and the radiation field
is depleted of photons with energies in excess of 13.6 eV.  In M16 there is a
clear distinction between the peak of the [S~II] zone and the peak of the
H$\alpha$ zone, but the two are separated by only of order 10$^{15}$ cm.  H96
found that this structure can be reproduced in detail by photoionization models
using a steeply falling density distribution.

Manipulation of photoionization models has revealed that the physical extent of
the emission zones and their relative strengths and separations can vary
substantially.  Factors that can affect these distributions include the
steepness of the particle density gradient toward the ionization front, the
intensity of the incident ionizing radiation, and the ``hardness'' of the
ionizing radiation (the fraction of photons that fall shortward of the Lyman
edge)(H96, and references therein).

While 30 Doradus is too distant to allow resolution of the separation between
the H$\alpha$ and [S~II] peaks, it is close enough to see the difference
between the highly localized [S~II] emission and the thicker zone in which much
of the H$\alpha$ and all of the [O~III] originates.

In 30 Doradus, our images have been taken in the three emission lines mentioned
above.  Comparing Figures 1-3 we notice that while the images differ in several
ways but are also quite similar in many others.  In the majority of locations
across the nebula the general appearance in the distribution of [O~III]
emission is almost identical to that of the H$\alpha$ emission.  The nebula
looks very different in [S~II] $\lambda\lambda$6717+6731 but since these lines
are brightest outside of the ionized volume, they trace structure that [O~III]
and H$\alpha$ cannot.

Most of the physical structures we observe in 30 Doradus exhibit bright
emission in all three of the lines allowing us to trace a consistent picture of
the ionization structure in those locations.  Several regions appear only
bright in H$\alpha$ and [S~II] indicating that the emission from those regions
is characteristically of lower excitation.

In the rest of this section we are going to visit and describe individual
locales in the wall of the nebula.  Each will be featured as a morphological
prototype for a host of other structures we observe across the field but that
we do not have the space to address individually here.  The approach we will
use employs the three emission lines to compare and contrast the observed
physical structure in each line and uses these clues to make some statement
about the physical nature of each object.

\subsection{Objects of Interest in the Field}

\subsubsection{A high proper motion star?}

Toward the upper left corner of Figures 1-4,6 is a long tube-like structure,
especially visible in [S~II], that runs approximately E-W.  It is rather dim
and so has been reproduced in some detail as Figure 7a-7c.  In Figures 7-9 the
sets of 3 panels observe the same convention: the top panel represents [O~III],
the middle represents H$\alpha$, and the bottom represents [S~II].

This object is composed of a long (1.5pc) limb-brightened tube of emission that
is bright in [S~II] and is visible in H$\alpha$ but not [O~III].  There is one
part of the tube that does show up in [O~III] (in the upper left) and this
corresponds to regions where the H$\alpha$ and [S~II] are brightest.  The wall
of the tube is also thickest in this region.  Elsewhere the wall is close to be
being unresolved, and appears broken and discontinuous in many locations.  The
lower right end of the tube appears to widen to a more spherical shape where a
faint star is seen at the heart of the cavity.  The very end of the cavity is
very fragmented in appearance, and has a radius of about $7\times10^{17}$ cm. 
The center volume of the tube and cavity are relatively dark contributing
little emission.  The low ionization tube lies within a more extended region of
diffuse higher ionization emission.  The intensity of the emission from the
tube does not depend on distance from the star at its head, indicating that it
is ionized externally rather than by the visible star.

The appearance of the system is one of a tube with a relatively dense wall,
that provides enough local electron density and column depth to make it
brighter in both H$\alpha$ and [S~II] relative to the background nebula.  The
walls appear to have a resolvable thickness in places but this appearance might
also be produced by limb brightening of an unresolved surface.  The apparent
``width'' of the wall in the upper left is about 1.3 WFC pixels or $10^{17}$cm.
We infer that the center of the tube is less dense because of the lower
emission in that region. The tube is seen to narrow with increasing distance
from the star.

We suggest that this structure might be the result of a high proper motion star
that has a stellar wind.  This star appears to be ploughing its way through a
cloud of gas of intermediate density.  Around the star itself the wind
establishes a balance with the thermal pressure of the local gas and produces
the observed cavity.  As the star moves through the cloud, from upper left to
lower right in our image, the leading edge of the cavity continues to be pushed
into the local gas.  On the trailing edge the wall of the cavity persists to
leave a channel or tube through the nebula gas.  

If we balance the thermal pressure in the main nebula, obtained from xray
observations (Wang \& Helfand 1991) to be about 10$^{-10}$ ergs cm$^{-3}$, with
the kinetic energy of the material of material in a wind, then we can infer the
density of the material in that wind.  The typical velocity of a stellar fast
wind is about 10$^{3}$ km s$^{-1}$, which implies a wind particle density of
about 10$^{-2}$ cm$^{-3}$.  This corresponds to a mass loss rate of about
10$^{-7}$ M$_{\odot}$ yr$^{-1}$ when integrated across a sphere of radius
$7\times10^{17}$ cm.  At the point where the wind kinetic energy balances the
thermal pressure in the nebula, ie. at the bubble wall, if the gas is at a
temperature of 10$^{4}$ K and using the known pressure, we calculate a local
particle density of about 50 cm$^{-3}$.

This is a fairly low density, and one would expect there to be significant
[O~III] and H$\alpha$ emission except that the surrounding material provides
enough shielding.  This calculation simply balances the momentum in the wind
with the thermal pressure of the local nebula gas.  If the star does indeed
have a significant proper motion then the mass loss rate and the gas density in
the wall are both lower limits.

This tube will collapse in time as the gas in the tube cools and the
surrounding nebula gas moves to fill the void under thermal pressure.  The
speed of this collapse would be something like the local sound speed.  Using
the canonical temperature for photoionized gas of 10$^{4}$K we estimate the
local sound speed to be of order 12 km s$^{-1}$.  We measure the opening
half-angle of the tube to be 3 degrees.  If all the transverse velocity is
produced by the collapse of the tube at the sound speed, the 3 degree angle
implies that the forward proper motion of the star is of order 230 km s$^{-1}$,
which is equivalent to an angular proper motion of 0.95 mas yr$^{-1}$.  

If this is a star with a substantial proper motion, then its path appears to
take it directly toward the center of the cluster.   If we take the mass of the
cluster to be 10$^{6}$ M$_{\odot}$ then at the apparent distance of the tube
from the cluster center of about 20 pc the escape velocity of the cluster is
about 20 km $s^{-1}$.  We have calculated an apparent proper motion that is an
order of magnitude greater than this.  Therefore the star is not bound, and is
therefore not on a return orbit.  The fact that its velocity vector appears to
coincide with the cluster is coincidence.

\subsubsection{Overlapping bubbles or a single cavity?}

Towards the center of WF2 in Figures 1-4,6 are a couple of overlapping arcs. 
The arcs are shown in Figure 7d-7f, and are visible to varying degrees in all
three emission lines.  The sharpest structure is seen in the [S~II] image. 
Both arcs are covered or masked by appreciable levels of foreground H$\alpha$
and [O~III] emission.  The arcs themselves appear quite sharp-edged, and appear
to brighten towards their edges suggesting that they are probably thin sheets
that are limb brightened, implying that they are seen in projection from one
side.  

At the distance we adopt to the nebula the arcs have a radius of about 1.5
parsecs with a peak surface brightness of 0.022 ergs cm$^{-2}$ s$^{-1}$
Sr$^{-1}$, or $5.1\times10^{-13}$ ergs cm$^{-2}$ s$^{-1}$ arcsec$^{-2}$.  

Each of the arcs appears to open toward one of the two bright stars seen in the
figure.  The lower of these two stars has been identified in the ground-based
census of stars in the region of the R136 cluster published by Parker 1993, and
subsequently included by Walborn \& Blades 1997.  The other star has not been
previously identified, but does appear from our data to be redder and therefore
later in spectral type than the first star.  The lower star is classified as a
B 0.5 type Ia star by Parker 1993, implying the existence of a possible massive
wind associated with it.  We therefore assume that the lower star is the one
responsible for the origin of the arcs, with the upper star being a line of
sight coincidence with the structures.  Recent work by Aufdenberg et al 1998
has shown that conventional atmosphere models of early B stars have
underestimated the hydrogen-ionizing flux they are capable of producing, in
some cases by as much as a factor of two. The helium-ionizing flux may be
underestimated by an order of magnitude.  As such it is entirely possible that
the lower of these two stars could produce enough flux to ionize the local gas
and cause a lot of the diffuse H$\alpha$ we see spread across the foreground of
this object.  The wind from the star might also have been sufficient to sweep
up enough material from the region around the star to produce the thin shell of
gas we observe.  This shell was then ionized by the flux from the same star.

\subsubsection{Finger-like column in the molecular wall of the cavity}

Immediately to the south of the R136 cluster is an isolated region of very
bright emission in all three emission lines we observed.  Figure 8a-8c
illustrate the structure in the previous manner.  Central to the overall
morphology of the region are a number of finger-like structures that stick out
from the wall of the cavity directly at the main cluster.  The main column is
about a parsec long and about 0.25 parsec wide.  The [O~III] and H$\alpha$
emission are again very similar, while the [S~II] is quite different being
concentrated to a narrow region that hugs the wall of cavity.  As such these
structures bear a striking resemblance to the columns of dense material
observed in M16, the Eagle Nebula, in our own Galaxy and recently studied in
high detail with photoionization modelling by H96.  The main column we observe
in 30 Doradus is directly comparable to the largest column in M16.

Considering the physical scales found to be important in the modelling of the
structure observed in M16, we know that these scales are hopelessly unresolved
in the 30 Doradus nebula at a distance some 25 times further away than M16. 
Direct comparison between the two nebulae also shows a difference in
presentation - with the Doradus finger being viewed more obliquely and through
much more foreground emission than M16.  Since the visible face of the column
is bright in both H$\alpha$ and [S~II] we can state that the near face of the
column is directly exposed to the radiation from the cluster.  As such we can
also state that the column is therefore on the far side of the tangent plane
that passes through the cluster, and that therefore the finger is sticking out
of the back wall of the nebula rather than a feature of the near wall.

We place the finger on the back wall but not too far into the back since the
foreground emission is not as bright as some regions in the nebula.  By doing
this we can infer very low levels of intrinsic reddening within the nebula, but
we will still adopt our conservative estimate of 1 magnitude of extinction in
A$_{V}$.

Based on the similarity in the morphology of the two systems, we explicitly
compare M16 with the Doradus finger in the Discussion section below and derive
limits on the physical conditions in the nebula despite our much poorer
physical resolution at the distance of the LMC.  For more details about the
structure of this object refer to that section.

\subsubsection{A low excitation H~II region}

At the bottom of the field in Figures 1-4,6 is a well-defined bubble-like
structure about 2-3 pc in diameter.  This structure is depicted in the usual
way in Figure 8d-8f.  It is characterized by bright emission in both H$\alpha$
and [S~II].  At the geometric center of the region is a bright star not
catalogued by any previous surveys of the region.  Around this star is some
very faint diffuse [O~III] emission.  

These morphological hallmarks when combined with the absolute levels of the
ratio of [S~II]/H$\alpha$ (ranging from less than 0.1 in the center to peaks
around 0.3 at the edge) and [O~III]/H$\alpha$ (only ranging between 0 and 0.4
across the bubble) point to a low-excitation H~II region centered on the star. 
This view is supported by the fact that the peak in the lower-energy emission
lines [S~II] occurs exterior to that of the higher line H$\alpha$ - a known
morphological hallmark of a photoionized region.  We observe the peak H$\alpha$
surface brightness to be 0.004 ergs cm$^{-2}$ s$^{-1}$ Sr$^{-1}$, or
$9.1\times10^{-14}$ ergs cm$^{-2}$ s$^{-1}$ arcsec$^{-2}$.

The lack of any appreciable [O~III] emission in the main body of the small H~II
region, particularly around the star itself, implies that the ionizing
radiation from the star is not very ``hard''.  However, there is still enough
flux coming out of the star to ionize an appreciably large volume of gas.  This
would suggest an early B star as the culprit rather than a late O star.  In
addition the lower left end of the bubble appears open-ended and may represent
some type of over-pressure blowout from the cavity into a lower density region.
Further observations would be needed to clarify the exact physical nature of
the system.

An excellent collection of Galactic H~II regions was catalogued by Sharpless
1953, chosen for their strong line emission.  Many objects from this sample
have been observed by Hester et al 1992 as part of a morphological atlas of
line emission from H~II regions.  After comparing this bubble-like object with
several of the nebula in this collection, the best analog found was the nebula
S-104.  The observed line ratios and overall shape were very similar.  We
assume that the two nebulae are also similar in the source and strength of
their ionizing sources, allowing us to assume that they are the same physical
size based on Str\"{o}mgren sphere considerations.  To make the observed
diameter of S-104 the same as that of the bubble nebula it would have to be
placed at a distance of about 50kpc.  This confirms that the bubble nebula is
located in the LMC.

It is difficult to place the bubble along the line of sight with respect to the
30 Doradus nebula.  Some of the structure observable in the 30 Doradus nebula
can be seen through the bubble, suggesting that the bubble might be closer to
the observer than the main nebula.  However, as we stated above, we do know
that the bubble nebula is part of the LMC and is therefore more local to 30
Doradus than to our Galaxy.

\subsubsection{Two views of the edge of the main cavity}

Examination of the images in Figures 1-4,6 show a remarkably circular edge to
the main cavity - that part of the inner nebula immediately centered on the
R136 cluster.  This region has a radius of about 8pc and in our field is
characterized by a bright rim to the SW with numerous large elephant-trunk-like
structures.  In Figure 9 we present two sets of panels focussing on two
particular areas around this bright rim.  

Figure 9a-9c depicts a region WSW of the cluster with a dark ridge that runs as
a chord from rim edge to rim edge.  Scattered along the outside of this ridge
are several blob-like stuctures with fairly long tails.  At this distance our
view of these objects is not good, but we can determine some morphological
properties.  The objects are bright in both H$\alpha$ and [O~III], with the
latter appearing marginally sharper around the edge of the objects.  In [S~II]
the objects appear smaller and fainter.  These structures are about 2-3 WFC
pixels across ($2\times10^{17}$cm, 13000 AU, 0.06 pc) and have tails as long as
10-12 pixels ($7-10\times10^{17}$cm, 65000 AU, 0.3 pc).  The distribution of
emission suggests that the [S~II] emission comes from a central core, with the
H$\alpha$ and [O~III] emission coming from a bright sheath surrounding the core.

As such these objects could represent smaller versions of the finger-like
column, mentioned above.  These structures could also be protruding from the
cavity wall or could even be separated from the wall. It is not uncommon to
find large globules of this kind floating free in Galactic H~II regions (we
have found examples in M8 and M17 from the collection of H~II regions by Hester
et al 1992), and so it should not be surprising to find analogs in this nebula.

A little further out from the cluster is the main inner cavity edge -- seen as
a large circular interface around the PC/WFC boundaries in Figures 1-4,6.  The
brightest part of this region is depicted in Figure 9d-9f.  In this field we
have two remarkable structures.  The lower, darker feature has many common
morphological traits with outflows observed in our own Galaxy associated with
HH objects and the like.  A good physical analog might be HH47 except that HH47
is only about 0.6 pc long (Heathcote et al 1996), whereas the ``flows'' we
observe in 30 Doradus are as long as 1.2 pc.  The flow is composed of dense,
non-optically-emitting gas and appears to broaden into a bow-shock-like
interface at its end.  The narrow end of the structure could be the outflow
source and appears to be buried in the cavity wall.

Immediately above this is another structure which appears quite ghostly in
Figure 9d-9f.  It is another tall column of material, but its shape and the
limb brightened emission in all three emission lines suggest that it is more
like the finger-like structure we considered above.  This column measures about
a parsec in length and so is very comparable in size to M16 and the Doradus
finger.  Buried in the head of this column is a stellar object that may or may
not be associated with the column.  This object is very close to one of the
bright IR sources found in the 30 Doradus nebula by Hyland et al 1992, and
noted in passing in WFPC-1 images of the region by Hunter et al 1995a.  The
accuracy of the position published by Hyland could place either of these two
column-like objects within the positional error bars.

The second column is veiled in very bright foreground emission making clear
definition of the structure hard.  When we compare the emission line profiles
across the edge of this structure, we see similar ionization structure
characteristics to those found across simple photoionized interfaces.  The
[O~III] and H$\alpha$ are more extended and appear to be limb brightened only
on the side of the column closest to the central cluster.  The [S~II] emission
is much more concentrated towards the very edge of the column and is spread
thinly across the face of the column as well.  Due to the bright foreground
emission in both [O~III] and H$\alpha$ we cannot accurately determine how
bright the front face of the column really is in these two lines.

Across this particular field we also see many smaller globules and finger-like
structures that exhibit similar physical morphology in the three emission lines
as do the prototypes we have already discussed.  In light of what was learned
about photoevaporation in M16 by H96, we expect the majority of the remaining
structures in the field to be somewhat denser than their surroundings, and
might harbor new stars that are in the process of forming.

The remarkable thing about the emission along this cavity edge is the total
lack of any emission to the upper left (toward the cluster) from this edge. 
This is apparent too in Figures 1-4,6.  The volume of the very inner cavity
appears totally devoid of any detectable line emission - a direct result of a
very low local electron density driving down the emission measure for these
lines.  The only line emission seen in this region is visible between the stars
of the cluster.  Estimates for particle densities in the main cavity of 30
Doradus based on diffuse xray background emission (Wang \& Helfand 1991)
suggest number densities around 0.2 cm$^{-3}$.  The same paper also noted
several regions or pockets of very hot xray emitting gas associated with
``holes" in the optical H$\alpha$ emission seen from the ground.  These regions
are probably filled with gas that has been even more strongly shock excited
than the majority of the cavity and may be associated with some of the
Wolf-Rayet stars discovered by Hunter et al 1995b.

\section{Discussion}

\subsection{H$\alpha$ Surface Brightness and Photoevaporative Flow}

In \S 2.2.3 we describe an elephant trunk structure that is a remarkable analog
to the structures in M16.  Located about 10$^{18}$ cm from the head of this
structure and another nearby elephant trunk are two arcuate emission features. 
We interpret these features as bow socks where the diverging photoevaporative
flow stagnates against the pressure in the hot gas that fills most of the
volume of 30 Doradus.  (A similar interpretation was offered by H96 for faint
filaments in M16).

Since the interface at the end of the column in 30 Doradus is concave to the
pillar, we assume that the flow is spherically divergent and obeys a 1/r$^{2}$
law for density with distance.  

We observe that the surface brightness of this concave bowshock is
$3\times10^{-4}$ ergs cm$^{-2}$ s$^{-1}$ Sr$^{-1}$ above the bright background,
which is equivalent to an emission measure of 1400 cm$^{-6}$ pc.  The observed
``width'' of the bowshock is very thin indeed.  It appears to be a
limb-brightened shell that has little or no transverse thickness.  We put an
upper limit on the width of the interface at 2 WFC pixels or 0.05 parsecs
($1.5\times10^{17}$cm).  If we assume we are looking at a tangent through a
shell of this thickness then the characteristic line of sight through the wall
of that shell will be of order 10$^{18}$ cm.  Using the emission measure we
observe for the shell wall this implies a local electron density of about 70
cm$^{-3}$.  This is a lower limit.  If this shell is in fact thinner than we
can resolve, which is quite likely, then the actual electron density will be
higher.  If the shell is a factor of 3 thinner (which is the sort of shell
thickness observed in Galactic wind blown bubble nebulae such as NGC 7635 and
NGC 6888) then the implied electron density is more like 260 cm$^{-3}$.  We
observe the radius of curvature of the bow shock to be about 24 pixels, with
the end of the column at a radius of about 7 pixels.  This implies a factor of
1/12 drop in the flow density from the end of the column to the bowshock,
assuming a 1/r$^{2}$ divergence.  Therefore the density at the end of the
column must be about 3000 cm$^{-3}$.

The thermal pressure at the end of the column drives the divergence of the gas
away from the interface. At the bowshock the pressure of the photoionized gas
is given by:

\begin{equation}
P = n k T_{e} = 2n_{e} k T_{e} = 2\cdot 260\cdot k \cdot 10^{4} \sim 7\times10^{-10} {\rm erg \>cm^{-3}}
\end{equation}

We need to determine the character of the material in the cavity of 30 Doradus.
If it is simply a volume filled with photoionized/photoevaporated material then
the typical electron temperature is something like $10^{4}$ K.  However, if the
interior gas has been shock heated by the strong stellar winds from the dozen
or so Wolf-Rayet stars found in the cluster (Hunter et al 1995b), combined with
the possible effects of supernovae that have occured in the cavity since it
was formed, then the typical electron temperature is probably in excess of
$10^{6}$ K.  Wang \& Helfand 1991 presented observations of the diffuse gas in
the nebula made with the {\it Einstein Observatory}.  They found a diffuse
x-ray component with a characteristic temperature of about $5\times10^{6}$ K. 
From their observations of the x-ray luminosity they presented a relation
between the electron density in the x-ray bright gas and the overall cavity
size.  In their paper they assumed that the photoionized gas pressure was more
like $1\times10^{-10}$ ergs cm$^{-3}$, instead of the higher value we
calculated by divergence above.

If we balance the divergent $\rho v^{2}$ flow calculated above with the thermal
pressure of an x-ray bright gas at $5\times10^{6}$ K, we predict an electron
density of 0.5 cm$^{-3}$ for the hot diffuse gas that fills most of the nebula.
This number compares well with the implied estimate by Wang \& Helfand of 0.2
cm$^{-3}$.  This is good agreement considering how disparate the two approaches
are and the assumptions that have been made.

Next we turn to the emission we see across the top of the column itself.  The
brightness of the emission at the end of the column allows us to make some
statement about the physical conditions in the region where the emission is
most intense.  Hunter et al 1995b quote an ionizing flux of lyman continuum
photons of $2\times10^{51}$ photons sec$^{-1}$ within 93 pc of the cluster
center.  This number was in good agreement with calculations made by Kennicutt
\& Hodge 1986 if an internal reddening of 1 magnitude in A$_{V}$ is assumed. 
Here we will adopt this value for the ionizing flux incident on the elephant
trunks since this close in to the center of the nebula the ionizing flux is
dominated by the radiation from R136.

If Q is the ionizing luminosity in units of photons sec$^{-1}$ coming out of
the cluster, then we can express the incident ionizing flux at the elephant
trunks as:

\begin{equation}
q = \frac{Q}{4 \pi R^{2}} = \frac{2\times10^{51}}{4\pi\cdot(5\times10^{19} {\rm cm})^{2}} = 6\times10^{10} {\rm photons \>cm^{-2} \>sec^{-1}}
\end{equation}

\noindent where R (= 15 pc) is the distance from the ionizing stars.  This
estimate is made in the plane of the sky and could be much larger if the finger
is located on the back wall of the nebula cavity. Assuming that objects such as
these are randomly distributed throughout the cavity we can assess this
uncertainty by introducing a factor of $\sqrt {2}$ in our estimate of the
distance from the ionizing source.  The fraction of these incoming photons that
produce a visible H$\alpha$ photon is given by the ratio of the recombination
coefficient for the H$\alpha$ transition to the total recombination rate for
hydrogen as a whole.  From Osterbrock 1989, Table 4.2, we find that the
recombination rate for H$\alpha$, $\alpha_{H\alpha}$ (in units of cm$^{-3}$
sec) is:

\begin{equation}
\alpha_{H\alpha} = \alpha_{H\beta}\cdot\frac{j_{H\alpha}}{j_{H\beta}}\cdot \frac{E_{H\beta}}{E_{H\alpha}} = (3.03\times10^{-14})\cdot(2.87)\cdot(1.35) = 1.17\times10^{-13} {\rm cm^{-3} \>sec}
\end{equation}

\noindent where E$_{H\alpha}$ is the photon energy for H$\alpha$, E$_{H\beta}$
is the photon energy for H$\beta$, and all numbers are quoted for conditions of
$10^{4}$ K.  This predicts that the fraction of H$\alpha$ photons emerging
versus incoming hydrogen ionizing photons is:

\begin{equation}
\Im = \frac{\alpha_{H\alpha}}{\alpha_{B}} = \frac{1.17\times10^{-13}}{2.59\times10^{-13}} = 0.453
\end{equation}

Using equations (2) and (4) we calculate that the expected H$\alpha$ flux
emerging from the end of the column should be $3\times10^{10}$ photons
cm$^{-2}$ sec$^{-1}$.  If we include the statistical factor of $\sqrt{2}$ for
the projection along the line of sight, this could fall as low as
$2\times10^{10}$ photons cm$^{-2}$ sec$^{-1}$.  This is the emission we expect
to be emerging from the end of the column, but we need to factor in the adopted
extinction to make an estimate of the expected surface brightness we would
measure.  Doing this, using 1 magnitude of A$_{V}$, we end up with an estimate
of $1\times10^{10}$ photons cm$^{-2}$ sec$^{-1}$.  

We need to estimate the size of the emitting region, but lack one of the
dimensions.  We can measure the transverse width across the column, and the
depth of the zone along the radial vector from the cluster, both measured in
the plane of the sky.  The third dimension, the transverse depth of the
emitting zone across the column but perpendicular to the plane of the sky, is
unknown.  Using our observations we will constrain this depth.  If we assess
what the emission per unit length across the top of the column is in the plane
of the sky, we can use the observed surface brightness to calculate how deep
the emission zone must be to produce that level of emission.

The first step is to place an aperture across the top of the column and
integrate it across the short dimension of the column.  This takes the measured
surface brightness in units of ergs cm$^{-2}$ s$^{-1}$ pixel$^{-1}$ and
produces an emission per unit length along the top of the column in units of
ergs cm$^{-2}$ s$^{-1}$ pixel$^{-1}$.  This is now essentially an integrated
profile of the emission across the interface at the end of the column.   From
this profile we need to choose the value and location that corresponds to the
thin emission zone at the top of the column, which we know to be unresolved in
our data.  When we do this we measure a value of $2\times10^{-14}$ ergs
cm$^{-2}$ s$^{-1}$ in the one pixel that represents the width of the emission
zone at the top of the column.  Converting this to H$\alpha$ photons, we get a
flux of $7\times10^{-3}$ H$\alpha$ photons cm$^{-2}$ s$^{-1}$ in one pixel. 
This flux needs to be corrected for the 1 magnitude of extinction we have
assumed, increasing the flux estimate to $2\times10^{-2}$ H$\alpha$ photons
cm$^{-2}$ s$^{-1}$ in one pixel.

We next need to take out the effect of the inverse square law by integrating
the flux we observe over a sphere of radius 51.3 kpc.  This yields a flux of
$5\times10^{44}$ H$\alpha$ photons s$^{-1}$ in that one pixel.  The width of
that one pixel is $7.7\times10^{16}$ cm at the distance we are assuming for 30
Doradus.  Dividing out the physical extent of the pixel along the top of the
column we end up with a flux per unit length of $6\times10^{27}$ H$\alpha$
photons cm$^{-1}$ s$^{-1}$.  Our estimate of the expected H$\alpha$ flux from
the top of the column was $1\times10^{10}$ photons cm$^{-2}$ sec$^{-1}$.  If we
divide the two quantities we arrive at an estimate for the depth of the
emitting region into the plane of the sky.  This depth is $6\times10^{17}$ cm,
or about 8 WFC pixels.  This dimension is very close to the observed transverse
width of the column in the plane of the sky, so it is appropriate to think of
the column as being cylindrical.

Now that we know the dimensions of the emitting region at the end of the
column, we can use the observed surface brightness to make an estimate of the
local electron density in the emitting region, and compare it to the estimate
of 3000 cm$^{-3}$ made independently using the observed bow shock around the
end of the column.

The conversion between observed surface brightness and emission measure (EM) is:
\begin{equation}
{\rm EM} ({\rm cm}^{-6} {\rm pc}) = 2.41\times10^{3} {\rm T}^{0.92} {\rm S}({\rm H}\alpha)
\end{equation}
where S is the surface brightness expressed in units of ergs cm$^{-2}$
s$^{-1}$ Sr$^{-1}$ (Peimbert et al 1975).

We measure the mean surface brightness per pixel to be $1\times10^{-13}$ ergs
cm$^{-2}$ s$^{-1}$ arcsec$^{-2}$, or $4\times10^{-3}$ ergs cm$^{-2}$ s$^{-1}$
Sr$^{-1}$. Correcting for an extinction of 1 magnitude raises these numbers to
$2\times10^{-13}$ ergs cm$^{-2}$ s$^{-1}$ arcsec$^{-2}$, or $1\times10^{-2}$
ergs cm$^{-2}$ s$^{-1}$ Sr$^{-1}$.  This measurement was made across the
brightest region of emission at the top of the column, after having the bright
background/foreground subtracted.  The corresponding emission measure is
$1\times10^{5}$ cm$^{-6}$ pc.  Using the derived path length through this
emission zone of $6\times10^{17}$cm, or 0.2 pc, we obtain an electron density
of about 700 cm$^{-3}$.

When considering the emission at the end of this column we have to assess the
impact of our lack of sufficient resolution.  We have calculated that if you
take the observed flux and the observed extent of the emitting region we derive
a mean electron density of about 700 cm$^{-3}$.  In M16, H96 shows that the
H$\alpha$ bright emission zone is of order $2\times10^{15}$ cm which would be
unresolved in our data.  Using the data from the HST observations of M16 we
estimate that the ratio between the observed mean electron density for the
region and the value derived from where the emission peaks is about a factor of
1/3.  We also have to assess the difference in path length through the emitting
gas.  The path length of the line of sight through the region where the
emission peaks will be shorter than the path length through the extended
emission (since it is further out from the interface).  We estimate this
difference in path length to be about a factor of 5 (or a factor of $\sqrt {5}$
in electron density since it is proportional to the square root of path
length).  Combining these two corrections yields a potential increase in our
electron density estimate to about 4700 cm$^{-3}$.  This is an average
assessment of the impact of the two effects, and it is entirely possible that
slightly different factors should be used in the case of 30 Doradus.   Given
this uncertainty this estimate compares well with our earlier independent
estimate of 3000 cm$^{-3}$ based on the bowshock.  A summary of the numbers we
have calculated is included in Table 2 along with numbers calculated for M16
from H96.

In Figure 10 we compare the finger in 30 Doradus with the columns of gas in
M16.  In addition we present measured profiles for the three emission lines to
illustrate the similarities between the two objects.  There are definite
differences between the two cases, however, which we will address below.

In conclusion, we have shown that the observed characteristics of the H$\alpha$
emission at the edge of the 30 Doradus nebula are well explained by the
existence of a photoevaporative flow.  We resolve a bowshock transition between
the photoevaporative flow and the shock heated tenuous gas that fills most of
the 30 Doradus cavity.  The apparent brightness of this interface gives us an
independent estimate of the conditions at the head of the column and agree well
with our estimates based purely on the apparent surface brightness coming from
the column itself.  The column in 30 Doradus is a good morphological analog for
the columns observed in M16 by others, but upon closer examination the
conditions in 30 Doradus are slightly different.

\subsection{Stratification of the Ionization Structure}

At the distance to the Large Magellanic Cloud (51.3 kpc) our linear resolution
in 30 Doradus is $7.7\times10^{16}$cm, or about 5000 AU.  This compares to the
typical resolution achieved by ground-based observations of Galactic H~II
regions of about $3\times10^{16}$ cm, or about 2000 AU.  With the HST we can
study the ionization structure in 30 Doradus in the same way that we have been
used to studying the structure in local objects.  

We cannot, of course, achieve the kind of resolution that was achieved with HST
imaging of M16 (H96) -- that allowed successful modelling of the observed
structure of the photoionized gas using the known incoming ionizing radiation
and the observed density profile of the ionized gas across the edge of the H~II
region.  It is, however, instructive to compare what information we can extract
from these data with the picture reached in M16.

The method used to model the structure in M16 started with the observed
H$\alpha$ profile.  For a constant path length through the emitting gas, the
surface brightness of H$\alpha$ emission is proportional to the square of the
local electron density.  This fact allows us to map the local density of
neutral gas, which in turn allows a self-consistent photoionization calculation
to be made of the observed structure using photoionization codes such as CLOUDY
(Ferland et al 1996).  The results achieved in M16 with HST were impressive and
represented the first time this type of model had been compared directly with
observation.  

To pursue this approach we need a clean view of an ionization front, and the
structure depicted in Figure 10 is again the best choice in our field.  We
imaged the nebula in three emission lines, each characteristic of a different
photoionization energy.  The structure of the ionization front in an H~II
region is predicted by theory to be layered with [O~III] emission originating
interior to the interface, the H$\alpha$ emission being concentrated toward the
interface and peaking there (by definition), and finally the [S~II] is
predicted to peak just outside the ionization front (since it has an ionization
potential just less than 13.6 eV) with an emitting zone that is very narrow.  

When profiles of sufficient resolution are taken across the bright edges of
H~II regions these emission line zones form ``strata'' where the emission from
each line forms a distinct peak.  These peaks are usually separated since the
conditions for the emission to be maximized for each line differ in terms of
the local electron density and the emissivity of the appropriate species.

In Figure 11a we show the background subtracted profiles from the column in 30
Doradus plotted as a function of linear distance.  The H$\alpha$ and [O~III]
peaks are essentially coincident.  The [S~II] peak is separated from the other
two by as much as $10^{17}$cm.  Employing the density distribution derived from
the square root of the H$\alpha$ profile in Figure 11a, we ran a model of the
expected photoionization structure at the interface between the H~II region
cavity and the bounding molecular gas.  The other input to this model is the
expected hydrogen-ionizing flux from the cluster (calculated above, Hunter et
al 1995b).  The appropriate mix of stellar continua was made using the spectral
type distribution inferred by the results of Hunter et al.  It should be noted
that the resulting photon distribution was {\it harder} than the continuum
employed for the M16 models in that it had a higher proportion of high energy
photons shortward of the Lyman edge (at 912 \AA).

Figure 11b depicts the results of this model convolved to the linear resolution
of our observations.  The n(H) plotted is the self-consistent solution from the
model, which took as input the profile from Figure 11a.  At lower radii there
are departures from the input profile due to the nature of the fit employed and
the boundary conditions placed on the model.  In the region where the H$\alpha$
emission is strong, the resulting n(H) closely tracks the observed profile.  

Comparison of Figures 11a and 11b reveals a pretty good match.  The H$\alpha$
and [O~III] peaks merge, as we observe, and the narrow [S~II] peak becomes more
separated from the H$\alpha$ peak.  Thus we have done a reasonable job of
reproducing what our observations show as far as the distribution of emission
in the three main lines.  It is unclear, however, how unique this solution is.

Figure 11c is a reproduction of Figure 10d - it depicts the same set of line
profiles that we have extracted in 30 Doradus, but this time from ground-based
observations of M16, which have been observed at essentially the same linear
resolution.  Figure 11d shows a set of profiles for M16 taken from the best fit
models used in H96, except that this is the model used for column I (their
paper presented models calculated for column II).  It is included for direct
comparison with Figure 11c.

What these 4 graphs show is the remarkable similarity between the structure we
observe in 30 Doradus and the structure we observe from the ground in M16.  We
have matched our observations in 30 Doradus with a model calculated using the
same approach as that used by H96.  Most of the morphological features are well
reproduced - peak separation, approximate zone width, ordering, and so on. 
However, the critical issue is that while this appears to be a good fit it has
been calculated using parameters that we know are incorrect - we know that the
actual width of the emitting zones are far narrower than we can resolve.  This
problem is well illustrated in Figures 11c and 11d.  From the ground we can
extract profiles that are very comparable to 30 Doradus and could calculate a
similar model in the same way.  However, from the model presented in Figure 11d
we know that the emitting zones for each emission line are very much narrower
than we can see from the ground.

Turning this around, it does allow us to make an important statement about the
structure we observe in 30 Doradus.  Throughout this section we have shown in a
variety of ways that a picture of a photoevaporative flow is consistent with
all the observations we have made.  The calculations we have performed by
several different methods have yielded the same physical parameters and
together the data support the case for the emission we observe in 30 Doradus to
originate in a photoionized photoevaporative flow.  By comparing Figures 11c
and 11d, and then comparing Figures 11b with 11a we can make the statement that
the structure we see in 30 Doradus is almost certainly composed of emitting
zones far narrower than we can resolve, but that appear the way we would expect
them to given the resolution we can achieve.  The structure and emission we
observe in 30 Doradus is well explained by exactly the kind of structure
observed with HST in M16.

There are differences between 30 Doradus and M16 but they are not great.  The
most significant difference between the two sets of profiles is that in 30
Doradus the H$\alpha$ and [O~III] peaks coincide whereas in M16 they are well
separated.  We have already stated that the ionizing flux is more intense in
M16, but this should not affect the separation of the H$\alpha$ and [O~III]
peaks very much.  What is different about the two objects is that the ionizing
continuum in 30 Doradus is much {\it harder} than it is in M16 due to the
presence of more very massive O stars and a handful of W-R stars.  The effect
of hardening the ionizing continuum is to compress the [O~III] zone toward the
ionization front since more photons of high enough energy will get through to
those regions.  This would, by inspection of Figure 11, shift the [O~III] peak
toward the right making it more coincident with the H$\alpha$ peak.  The
picture is still consistent.

\section{Conclusions}

We have presented high resolution narrow-band imagery of the 30 Doradus nebula.
There are many interesting localized structures within the nebula, a number of
which appear to be associated with winds and UV from stars that are not part of
the main 30 Doradus cluster.  However the majority of the emission from the
nebula is due to photoionization by the flux from the central cluster.  This
emission is largely concentrated in thin regions located at the interface
between dense molecular material and the shock-heated interior of 30 Dor.

At the resolution of the {\it HST} data we find that the structure in 30
Doradus is remarkably similar to what is seen in ground-based observations of
nearby H~II regions.  This similarity is not surprising given that despite an
overall difference in scale, locally the physical conditions in 30 Doradus are
not much different than those found in smaller H~II regions.  We demonstrate
this point above by focussing on one particular region in 30 Doradus and
showing that it is a very direct analog of the Galactic H~II region M~16. 
Taking this same argument a step further we are lead to the conclusion that
underlying the observed structure in M~16 is the same sort of extremely
localized and sharply stratified structure seen in the {\it HST} images of
M~16.  Thus, even though the 30 Doradus nebula spans hundreds of parsecs, the
emission from this giant H~II region arises largely in the same sorts of
sharply stratified photoionized photoevaporative flows seen in nearby H~II
regions.

The 30 Doradus nebula is a crucial case.  The fact that at {\it HST} resolution
30 Doradus is so similar to ground based images of nearby H~II regions has
allowed us to bootstrap our physical understanding based on detailed study of
nearby regions into the physical context of a giant H~II region surrounding a
massive young cluster.  Similarly, preliminary analysis of {\it HST} images of
more distant giant H~II regions suggests that they compare favorably with 30
Dor when that nebula is viewed at the same physical resolution.  This indicates
that conditions in 30 Doradus are probably typical of those found in these
distant H~II regions as well.  Bootstrapping first from nearby H~II regions to
30 Doradus in this paper, and we anticipate from 30 Doradus to more distant
regions in later work, we are approaching the conclusion that the emission from
giant H~II regions megaparsecs distant is determined by the physics of
photoevaporative flows in which relevant physical scales can be as small as 100
AU or less.

The significance of this work lies in the conclusion that the detailed study of
nearby, well-resolved H~II regions is directly applicable to distant giant H~II
regions in much the same way that an understanding of radiative shocks that is
tested in nearby supernova remnants can be applied in a variety of contexts in
which the shock itself is not resolved.  Viewed from a different perspective,
interpretation of observations of distant giant H~II regions must take into
account the fact that much of this emission arises not in vast expanses of
ionized or even clumpy gas, but instead in well defined and highly stratified
photoevaporative flows localized to the surfaces of molecular clouds.

\acknowledgments{This work was supported by NASA grant NAS5-1661 to the WF/PC
IDT and NASA contract NAS7-1260 to the WFPC2 IDT.  This work was supported at
ASU by NASA/JPL contracts 959289 and 959329 and Caltech contract PC064528.}

\newpage

\section{Tables and Figures}

\begin{table}[htpb]
\caption{Details of observations made of 30 Doradus}
\vskip 0.2in
\begin{tabular}{lcrc}
\hline
Image Root Numbers &	Filter &	Exp. Time &	Date Taken \\
\hline
u25y0101t, u25y0102t, &	F336W &		10.0s &		2 Jan 1994 \\[-1.5ex]
u25y0103t, u25y0104t & &		100.0s \\
u25y0105t, u25y0106t, &	F555W & 	4.0s &		2 Jan 1994 \\[-1.5ex]
u25y0107t, u25y0108t, & &		40.0s \\[-1.5ex]
u25y0109t, u25y010at & &		200.0s \\
u25y0201t, u25y0202t, &	F547M & 	5.0s &		2 Jan 1994 \\[-1.5ex]
u25y0203t, u25y0204t & &		100.0s \\
u25y0205t, u25y0206t, &	F814W &		5.0s &		2 Jan 1994 \\[-1.5ex]
u25y0207t, u25y0208t & &		100.0s \\
u25y0209t, u25y020at, &	F170W &		10.0s &		2 Jan 1994 \\[-1.5ex]
u25y020bt, u25y020ct & &		100.0s \\
u25y0301t, u25y0302t &	F502N &		500.0s &	2 Jan 1994 \\
u25y0303t, u25y0304t &	F656N &		500.0s &	2 Jan 1994 \\
u25y0305t, u25y0306t &	F673N &		500.0s &	2 Jan 1994 \\
\hline
\end{tabular}
\end{table}

\newpage

\begin{table}[htpb]
\caption{\baselineskip 3ex Table comparing the derived properties for the
photoevaporative flow from the column in 30 Doradus with published numbers 
from H96.}
\vskip 0.2in
\begin{tabular}{lcc}
\hline
Parameter &	M16 (HST)&	30 Doradus\\
\hline
Dimensions of Column (pc) &
		$0.3\times1.03$ &	$0.2\times1$ \\
Incident Ionizing Flux (photons cm$^{-2}$ sec$^{-1}$) &
		$4\times10^{11}$ &	$6\times10^{10}$ \\
Total Ionizing Flux in Cluster (photons sec$^{-1}$) &
		$2\times10^{50}$ &	$2\times10^{51}$ \\
Mean H$\alpha$ Surface Brightness (ergs cm$^{-2}$ sec$^{-1}$ Sr$^{-1}$) &
		$6\times10^{-2}$ &	$1\times10^{-2}$ \\
Depth of Low Resolution H$\alpha$ Emission Zone (cm) &
		$5\times10^{17}$ &	$6\times10^{17}$ \\
Depth of High Resolution H$\alpha$ Emission Zone (cm) &
		$6\times10^{16}$ &	unresolved \\
Mean Electron Density in H$\alpha$ Bright Zone (cm$^{-3}$) &
		4000 &			3000 \\
\hline
\end{tabular}
\end{table}

\newpage

\begin{figure}
\caption{Mosaic of the WFPC-2 field of the center of 30 Doradus, seen in
H$\alpha$ $\lambda$6563 emission.  The field orientation is indicated.  The
core of the cluster and in particular the cluster of stars R136 are centered on
the PC.  The field is characterized by regions of very bright emission from
thin interfaces, with a broad diffuse component that covers the face of the
nebula. The names of the four WFPC-2 chips are included for the unfamiliar
reader.}
\end{figure}

\begin{figure}
\caption{Mosaic of the WFPC-2 field of the center of 30 Doradus, seen in
[O~III] $\lambda$5007 emission.  The field orientation is indicated.  The core
of the cluster and in particular the cluster of stars R136 are centered on the
PC.  The [O~III] emission follows the H$\alpha$ emission very closely and only
departs from it in subtle places.}
\end{figure}

\begin{figure}
\caption{Mosaic of the WFPC-2 field of the center of 30 Doradus, seen in [S~II]
$\lambda\lambda$6717+6731 emission.  The field orientation is indicated.  The
core of the cluster and in particular the cluster of stars R136 are centered on
the PC.  Since the [S~II] emission is confined to the regions just exterior to
the ionization fronts at the walls of the cavity, only the edges of the nebula
show up bright in these lines.}
\end{figure}

\begin{figure}
\caption{Mosaic of the WFPC-2 field of the center of 30 Doradus, seen in broad
band F547M emission.  Note the lack of any line emission from the intense
nebula.  This frame is used to continuum subtract the line images.  The field
orientation is indicated.  The core of the cluster and in particular the
cluster of stars R136 are centered on the PC.  Many stars are seen in the
nebula with some small amount of dust scattering evident in the regions of
brightest line emission.}
\end{figure}

\begin{figure}
\caption{CTIO CCD Image of the 30 Doradus Nebula.  The WFPC-2 FOV presented in
this paper is indicated at the heart of the object.  Note the large size of the
overall ionized volume.  North is up.}
\end{figure}

\begin{figure}
\caption{Three-color image of the 30 Doradus Nebula.  This is a composite of
four sets of data: in the nebula emission blue being [O~III] $\lambda$5007,
green being H$\alpha$, and red being [S~II]$\lambda\lambda$6717+6731; while the
stars are replaced with the F170W images as blue.  Scale and orientation are
indicated.}
\end{figure}

\begin{figure}
\caption{Tube-like structure found to the SE of the main cluster.  (a) View
seen in [O~III], (b) view in H$\alpha$, (c) view in [S~II].  Double arcs
located SSE of the main cluster.   (d) View seen in [O~III], (e) view in
H$\alpha$, (f) view in [S~II].}
\end{figure}

\begin{figure}
\caption{Finger-like structure found directly south of the main cluster, and
discussed at length in the text. (a) View seen in [O~III], (b) view in
H$\alpha$, (c) view in [S~II].  Low excitation H~II region located SSW of the
main cluster, that appears to be a foreground object unrelated to the main 30
Doradus cavity.  (d) View seen in [O~III], (e) view in H$\alpha$, (f) view in
[S~II].}
\end{figure}

\begin{figure}
\caption{Smaller blobs of material seen in emission against the bright nebula. 
These are located close to the cluster and the inner rim of the cavity. (a)
View seen in [O~III], (b) view in H$\alpha$, (c) view in [S~II].   Close-up
view of main rim in inner cavity.  Several outflow-like structures are visible.
(d) View seen in [O~III], (e) view in H$\alpha$, (f) view in [S~II].  }
\end{figure}

\begin{figure}
\caption{(a) Picture of the region south of R136 selected to perform ionization
structure analysis.  The bowshock used in the text is indicated on this
picture, along with another one just to the NE of it.  An arrow indicates the
direction the cut illustrated in (c) has been taken.  (b)  The most obvious
physical analog to (a) is M16 in our Galaxy.  This ground-based image of the
Eagle Nebula shows the same finger-like morphology and provides a direct
comparison for the derived ionization structure from 30 Doradus.  Note the
physical scales for (a) and (b) are the same, but the angular scales are very
different.  Overlaid are the extracted apertures for each dataset.  An arrow
indicates the direction the cut illustrated in (d) has been taken.  (c) Derived
emission line profiles for the aperture indicated on (a).  Note the broad tails
on most of the profiles.  (d) Derived emission line profiles for the aperture
shown on the M16 image.}
\end{figure}

\begin{figure}
\caption{(a) Extracted emission line profiles for the aperture shown in Fig. 10
in 30 Doradus.  The H$\alpha$ and [O~III] peaks coincide, while the [S~II] is
markedly separated.  The physical scale is indicated in the legend.  (b) Model
of the ionization structure in 30 Doradus using the n(H) from (a) and convolved
to the same linear resolution as (a). The peak separations are reproduced as
well as the coincidence of the overall profiles between H$\alpha$ and [O~III].
(c)  Reproduction of Figure 10(d) showing the ground-based resolution profiles
- note how similar they are in width to those seen in 30 Doradus.  (d) A model
of the structure observed in M16 by H96.  This shows how the ground-based data
does not represent the true width of the emitting zones.  This is also what is
happening in 30 Doradus.}
\end{figure}

\end{document}